\documentclass[twocolumn,aps,prb,amsmath,amssymb]{revtex4}

\pdfoutput=1

\usepackage{graphicx}% Include figure files
\usepackage{dcolumn}% Align table columns on decimal point
\usepackage{bm}% bold math
\usepackage{times}
\newcommand{\infig}[3]{\includegraphics{#1}}

\newcommand{\speed}{\ensuremath{v}}
\newcommand{\fermik}{\ensuremath{k_{\rm F}}}

\newcommand{\gate}{\ensuremath{V_{\rm G}}}
\newcommand{\condave}{\langle G \rangle_n }

\newcommand{\conddrude}{\ensuremath{g_{\rm D}}}
\newcommand{\condwl}{\ensuremath{g_{\rm WL}}}

\newcommand{\condwlb}{\ensuremath{\Delta g_{\rm WL}}}
\newcommand{\aspect}{\frac{W}{L}}
\newcommand{\aspecti}{\frac{L}{W}}

\newcommand{\meanfreetime}{\tau_m}
\newcommand{\riph}{Z}
\newcommand{\ripl}{R}
\newcommand{\randomfield}{\delta B_\perp}

\begin{document}

\title{Rippled Graphene in an In-Plane Magnetic Field: Effects of a Random Vector Potential}

\author{Mark B. Lundeberg}
\thanks{to whom correspondence should be addressed: {\it mbl@phas.ubc.ca}}
\author{Joshua A. Folk}
\affiliation{Department of Physics and Astronomy, University of British Columbia, Vancouver, BC V6T
1Z4, Canada}

{ \abstract
We report measurements of the effects of a random vector potential generated by applying an in-plane magnetic field to a graphene flake. Magnetic flux through the ripples cause orbital effects:
phase-coherent weak localization is suppressed, while quasi-random Lorentz forces lead to anisotropic magnetoresistance.
Distinct signatures of these two effects enable an independent estimation of the ripple amplitude and correlation length.
}

%\pacs{
%72.10.Fk 	% Scattering by point defects, dislocations, surfaces, and other imperfections (including Kondo effect)
%  %72.20.My 	% Galvanomagnetic and other magnetotransport effects (3D bulk semiconductors)
%72.80.Vp 	% Electronic transport in graphene 
%73.20.Fz %	Weak or Anderson localization 
%  %73.23.-b %  Electronic transport in mesoscopic systems -- No. Too general.
%  %73.23.Ad %	Ballistic transport (see also 75.47.Jn Ballistic magnetoresistance in magnetic properties and materials)
%  %73.43.Qt %	Magnetoresistance (in Quantum hall effect)
%  %73.50.Bk %	General theory, scattering mechanisms in thin films
%73.50.Jt %	Galvanomagnetic and other magnetotransport effects (including thermomagnetic effects) in thin films
%  %73.61.Wp %	Fullerenes and related materials
%  %75.47.Jn %	Ballistic magnetoresistance  -- No. This is hard drive read heads.
%  %75.70.-i 	% Magnetic properties of thin films, surfaces, and interfaces
%  %75.75.-c 	% Magnetic properties of nanostructures
%}

% Include the date command, but leave its argument blank.

\date{\today}

\maketitle

Graphene is a one-atom-thick carbon sheet with unusual electronic properties due to its two-dimensional honeycomb lattice\cite{kim,novoselov}.
As an ultrathin membrane, graphene easily wrinkles into the third dimension, with nanometer-scale ripples that have been observed in microscopic studies\cite{geimripples,nanolettripples,geringerripples,tikhonenko}. Strains associated with rippling are expected to modify electronic transport by generating random scalar and vector potentials, which would affect transport by suppressing anti-localization\cite{geimwl} and increasing the scattering rate\cite{2008RSPTA.366..195K}. Until this time, however, no transport measurements have directly probed graphene's ripples.

\begin{figure}[t]
\begin{center}
\infig{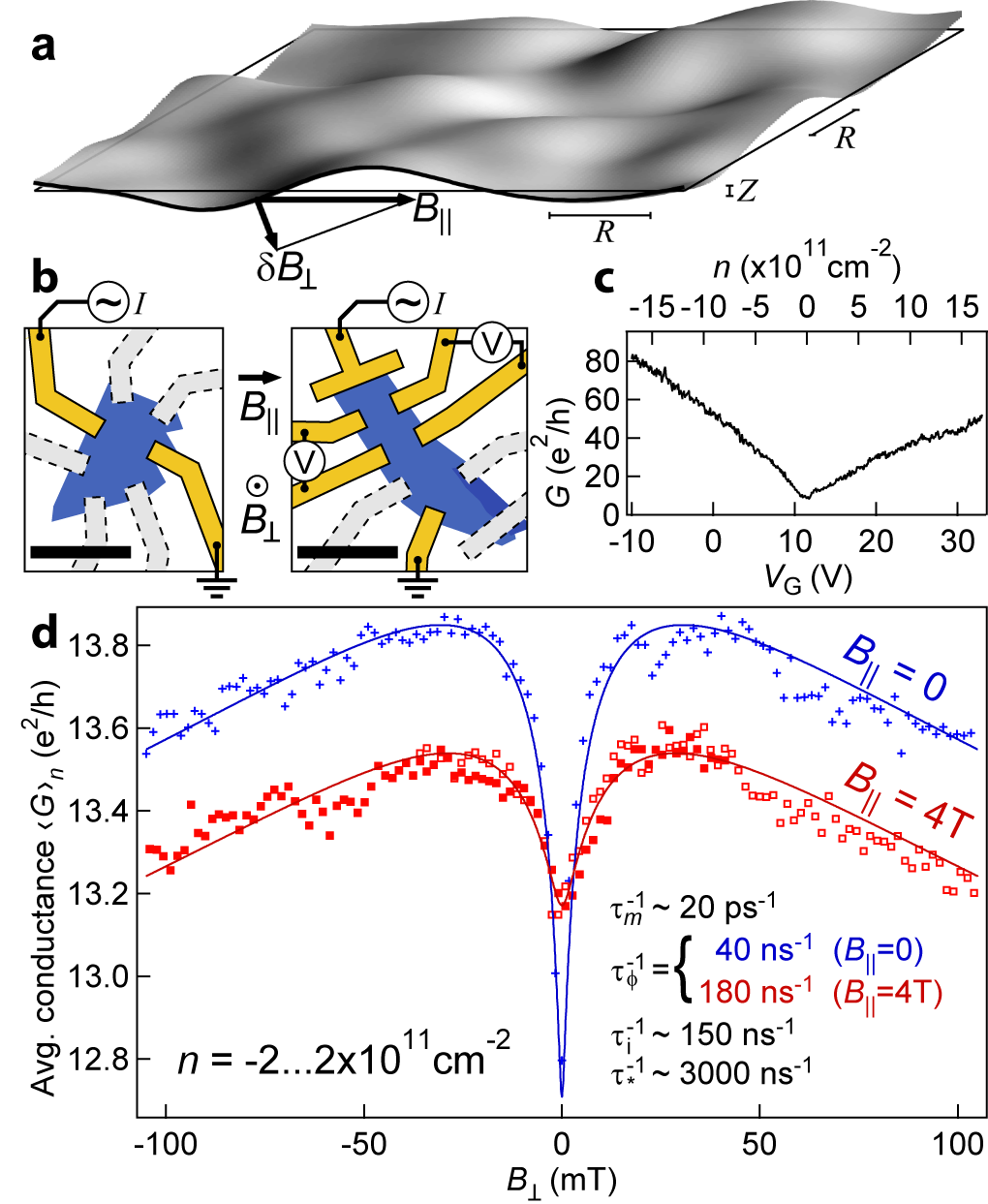}{245}{290}
\end{center}
\caption{
({\bf a}) Simulation of a rippled graphene sheet with a correlation length, $\ripl$, ten times its rms height, $\riph$.
The uniform in-plane field $B_\parallel$, applied to the rippled topography of graphene, leads to a random surface-normal field $\randomfield$.
({\bf b}) Schematics of graphene devices A (left) and B (right), showing orientation of applied fields $B_\parallel$, $B_\perp$, and electrical measurement setups (two-probe for A, multi-probe for B). Unused/broken electrodes are indicated by dashed edges. Scale bars are 5~$\mu$m.
({\bf c}) Conductance $G(n)$ for $B$=0, flake B.
({\bf d}) Weak-localization $B_\perp$-magnetoconductance of flake B at low density, measured at different in-plane fields: zero ($+$), $-4$~T ($\square$), $+4$~T ($\blacksquare$). Two fits to \eqref{wlb} are shown as solid lines. Applying the in-plane field dulls the central dip and decreases overall conductance. These are attributed to dephasing and scattering by the random $\delta B_\perp$.
} \label{fig1}
\end{figure}

Here, we report a magnetotransport measurement of graphene ripples using an in-plane magnetic field. In general, in-plane fields do not affect electron transport directly because electronic motion couples only to the component of magnetic field perpendicular to the graphene sheet. As illustrated in Fig.~1, however, a magnetic field aligned in the {\em average} plane of a graphene sheet ($B_\parallel$) includes an inhomogeneous perpendicular component ($\delta B_\perp$), which depends  on the local slope of the graphene flake. Ripples allow the in-plane field to affect transport directly by converting it to an inhomogeneous out-of-plane magnetic field, a form of random vector potental (RVP).

We observe two distinct effects of the $B_\parallel$-controlled RVP that each depend on the ripples' rms height, $\riph$, and correlation length, $\ripl$.
Random Aharonov-Bohm phases break time-reversal symmetry and suppress weak localization\cite{menszAlGaAs,dephasSiP,dephasInGaAs} (WL) with an effective dephasing rate\cite{mathurbaranger} proportional to $B_\parallel^2\riph^2\ripl$.
Random Lorentz forces\cite{geimrmf} lead to anisotropic momentum scattering by the in-plane field\cite{PhysRevB.70.193313} and anisotropic magnetoresistance $\propto B_\parallel^2\riph^2/\ripl$.
The effects are distinguished by measuring conductance in an additional {\em uniform} perpendicular field ($B_\perp$)---an independent experimental knob that can be used to suppress time-reversal symmetry.

Figure~1b shows a schematic of the graphene flakes and measurement set-up. Two flakes (called A and B) were prepared on SiO$_2$, electrically contacted, then cooled to an electron temperature of 40~mK. A two-magnet system provided independent control over the uniform magnetic field components $B_\parallel$ and $B_\perp$. The carrier density $n$ of the graphene was controlled capacitively using the Si back-gate (not shown in Fig.~1b), and the electrical conductance $G(n)$ showed a typical density-dependence (Fig.~1c) with mesoscopic fluctuations. The conductance fluctuations of flakes A and B showed spin-split character when an in-plane field was applied, as reported in our earlier work\cite{spinucf}. Here, we instead report the magnetic field dependence of the {\em mean} conductance $\condave$, averaging out (over intervals in $n$) the conductance fluctuations, following the procedure of Ref.~\onlinecite{tikhonenko}.

Figure 1d shows typical $\condave(B_\perp)$ curves for $B_\parallel=0$ as well as $B_\parallel=4$.  We first discuss the conventional case, $B_\parallel=0$.   The conductance can be separated into  Drude and WL contributions, $\condave(B_\perp)=\aspect [ \conddrude + \condwl(B_\perp) ]$, with device aspect ratio $\aspect$. The Drude conductivity, $\conddrude = 2 e^2 \speed\meanfreetime\sqrt{\pi|n|}/h$, depends on the momentum scattering time $\meanfreetime$ and Fermi speed $\speed \approx 10^6$~m/s and does not change significantly with $B_\perp$ for the fields used in this experiment. The $B_\perp$-magnetoconductivity,
$\aspecti[\condave(B_\perp) - \condave(0)]$,
is thus determined entirely by the WL component $\condwlb(B_\perp)$.

Weak localization is a phase-coherent back-scattering effect originating from the interference of a closed path with its reversed counterpart.   Paths that contribute to WL are limited in size by accumulated phases that are not symmetric under time (and path) reversal, coming from inelastic scattering, magnetic fields and some types of disorder.   Graphene's chiral charge carriers are intrinsically anti-localized, so $\condwl$ is positive as long as chiral memory is maintained along a path.  When chiral memory is lost due to intervalley scattering, however, conventional weak localization decreases $\condwl$.  Taking these effects together, a non-monotonic magnetoconductance curve is typically observed: large $B_\perp$ probes only short paths where chiral memory is retained, giving $\condwl>0$, whereas small $B_\perp$ also includes the contribution of longer paths for which chiral memory is lost, giving a conductance dip at $B_\perp=0$.

For 2D diffusive graphene, the WL contribution has been calculated to be:\cite{mccannwl}
\begin{align}
\condwlb(B_\perp)
 &=  \frac{e^2}{\pi h}
\bigg[F\Big(\frac{\tau_B^{-1}}{\tau_\phi^{-1}}\Big) 
- F\Big(\frac{\tau_B^{-1}}{\tau_\phi^{-1}+2\tau_i^{-1}}\Big) \nonumber \\
&\qquad\qquad{}- 2 F\Big(\frac{\tau_B^{-1}}{\tau_\phi^{-1}+\tau_i^{-1}+\tau_*^{-1}}\Big)\bigg],
\label{wlb}
\end{align}
where $F(z) = \ln(z) + \psi(\frac{1}{z} + \frac{1}{2})$ for digamma function $\psi(x)$.
Equation \eqref{wlb} depends on four rates $\{\tau_B^{-1},\tau_\phi^{-1},\tau_i^{-1},\tau_*^{-1}\}$ characterizing different mechanisms that suppress WL. The diffusive accumulation of Aharonov-Bohm phases from uniform $B_\perp$ gives $\tau_B^{-1} = 2\speed^2\meanfreetime e B_\perp/\hbar$.
The dephasing ($\tau_\phi^{-1}$), inter-valley scattering ($\tau_i^{-1}$), and intra-valley scattering ($\tau_*^{-1}$) rates each originate from scattering processes that each break different time-reversal symmetries.\cite{mccannwl}

The WL scattering rates were extracted from measured $\condave(B_\perp)$ curves by fitting to Eq.~\ref{wlb}, and are listed in Table \ref{fittable} for the $B_\parallel=0$ case. Values of $\meanfreetime$ (used to scale $\tau_B^{-1}$), as computed from $\conddrude \approx \aspect\condave$, were the primary systematic error since $\aspect$ was difficult to determine (see Methods section).
The longest time-scale, $\tau_\phi$, may have been saturated by the device dimensions at high doping, with $L_\phi = \speed\sqrt{\meanfreetime\tau_\phi/2} \sim 2\,\mu$m. The other characteristic lengths $L_i \sim 600$~nm, $L_* \sim 100$~nm, and $\speed \tau_m < 100$~nm, were not influenced by the sample geometry.

\begin{table}
\begin{tabular}{c c cccc}
\hline\hline
\multicolumn{2}{l|}{Parameter} & \multicolumn{3}{c|}{Flake B} & Flake A \\ 
&\multicolumn{1}{c|}{Units} & Hole & Low density & \multicolumn{1}{c|}{Electron} & Hole \\
\hline
$n$ &$10^{11}/$cm$^{2}$ & -13...-5 & -2...2 & 5...13 & -5...-3 \\
$W/L$ & - & $2.4\pm0.8$ & $2.0\pm0.7$ & $1.6\pm0.6$ & $0.7\pm0.3$ \\
$\meanfreetime^{-1}$ & $10^{12}/$s & $15\pm 5$ & $20\pm 10$ & $15\pm 5$ & $11\pm 5$ \\
$\tau_\phi^{-1}$ & $10^9/$s & $11\pm 1$ & $35\pm 8$ & $11\pm 1$ & $11\pm 2$ \\
$\tau_i^{-1}$ &$10^9/$s & $70\pm 50$ & $170 \pm 70$ & $120 \pm 80$ & $20 \pm 10$ \\
$\tau_*^{-1}$ &$10^{12}/$s & $5.3\pm 0.4$ & $2.7\pm 0.5$ & $2.1\pm 0.4$ & $4.0\pm 0.3$ \\
\hline\hline
\end{tabular}
\caption{Rates extracted from $\condave(B_\perp, B_\parallel = 0)$ at 40~mK (e.g. Fig.~1d);
$\condave$ was averaged over the specified density ranges. 
}\label{fittable}
\end{table}

Adding an in-plane field, $B_\parallel=4$T, changed $\condave(B_\perp)$ in two distinct ways (Fig.~1d).
First, the dephasing rate was increased, visible as a suppression of the WL dip at small $B_\perp$.
Second, the overall (Drude) conductance was reduced, causing an overall downward shift in the conductance at large $B_\perp$.
These effects can both be attributed to the ripple-induced RVP, and are discussed in order.

\begin{figure}[t]
\begin{center}
\infig{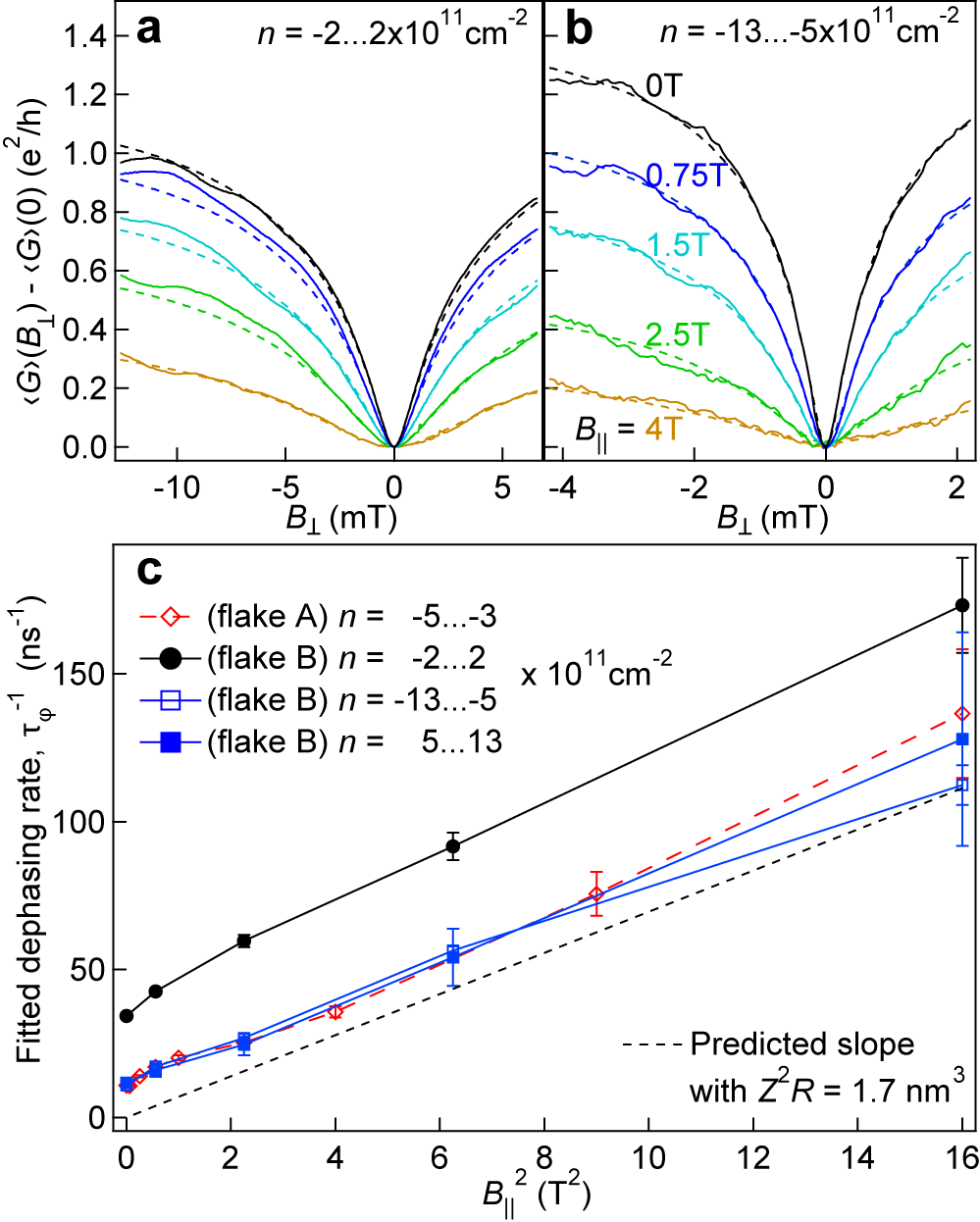}{245}{302}
\end{center}
\caption{
({\bf a,b})
$B_\perp$-magnetoconductance for small $B_\perp$, at various values of $B_\parallel$. Fits to Eq.~\ref{wlb} were computed
assuming $\tau_i$, $\tau_*$, and $\meanfreetime$ are independent of $B_\parallel$ (as in Table~\ref{fittable}),
while $\tau_\phi^{-1}$ was a free parameter.
(a) and (b) correspond to the low density (Fig.~1d) and hole-doped regions, respectively.
({\bf c}) For both flakes, extracted values of $\tau_\phi^{-1}$ increase in proportion with $B_\parallel^2$,
as predicted from Eq.~\ref{mbphi}.
Error bars plotted here do not include the larger systematic error from $\aspect$.
} \label{fig2}
\end{figure}

The additional dephasing effect of an in-plane field due to Gaussian-correlated ripples was calculated in Ref. \onlinecite{mathurbaranger}. Whereas a uniform $B_\perp$ affects WL through the diffusive rate $\tau_{B}^{-1}$, the RVP affects WL as a micro-scattering rate ($\tau_\phi^{-1}$) since the ripples are uncorrelated beyond short distances ($\ripl \ll \speed \meanfreetime \sim 100$~nm):
\begin{equation} 
 \tau_\phi^{-1} \rightarrow \tau_\phi^{-1} +
\sqrt{\pi} (e^2/\hbar^2) \speed \riph^2 \ripl   B_\parallel^2 .
\label{mbphi}
\end{equation}
Eq.~\eqref{wlb} was fit to multiple $\condave(B_\perp)$ curves at finite $B_\parallel$, allowing only $\tau_\phi^{-1}$ to change from the $B_\parallel=0$ fits (Fig.~2a,b), in order to extract the $B_\parallel$ effect.
Figure 2c confirms the $\Delta\tau_\phi^{-1}(B_\parallel) \propto B_\parallel^2$ dependence in Eq.~\ref{mbphi}, with a density-independent $\riph^2 \ripl = 1.7 \pm 0.5$~nm$^3$ extracted for flakes A and B (Fig.~2); the uncertainty is dominated by uncertainty in $\aspect$.
Although the WL rate $\tau_\phi^{-1}$ is commonly associated with inelastic scattering and loss of phase information, in this case it is enhanced by elastic scattering from the RVP which only scrambles phase information deterministically. The distinction can be seen in conductance fluctuations, which are softened by decoherence but only scrambled by elastic scattering. As reported in our previous work on flakes A and B, fluctuations were scrambled by $B_\parallel$, and only decreased by $\frac{1}{2}$ in variance due to broken spin symmetry.\cite{spinucf}

Besides suppressing localization, $B_\parallel$ caused an overall downward shift in the $B_\perp$-magnetoconductance trace away from zero field (Fig.~1d).  This shift indicates a change in the Drude conductivity; it was isolated from the dephasing effect by examining the $B_\parallel$-magnetoconductance for values of $|B_\perp|>50$~mT (Fig.~3), where the total WL correction\cite{mccannwl} $\condwl$ is essentially unaffected by the changes in $\tau_\phi^{-1}$.
The effect of RVP on Drude conductivity can understood as a decrease in $\meanfreetime$ due to scattering by the Lorentz forces from the RVP (Fig.~3a). A similar random-field resistivity has been observed in 2D systems subject to RVPs originating from nearby magnetic particles or superconducting vortices\cite{geimrmf,PhysRevB.70.193313, Wada20101138}.

\begin{figure}
\begin{center}
\infig{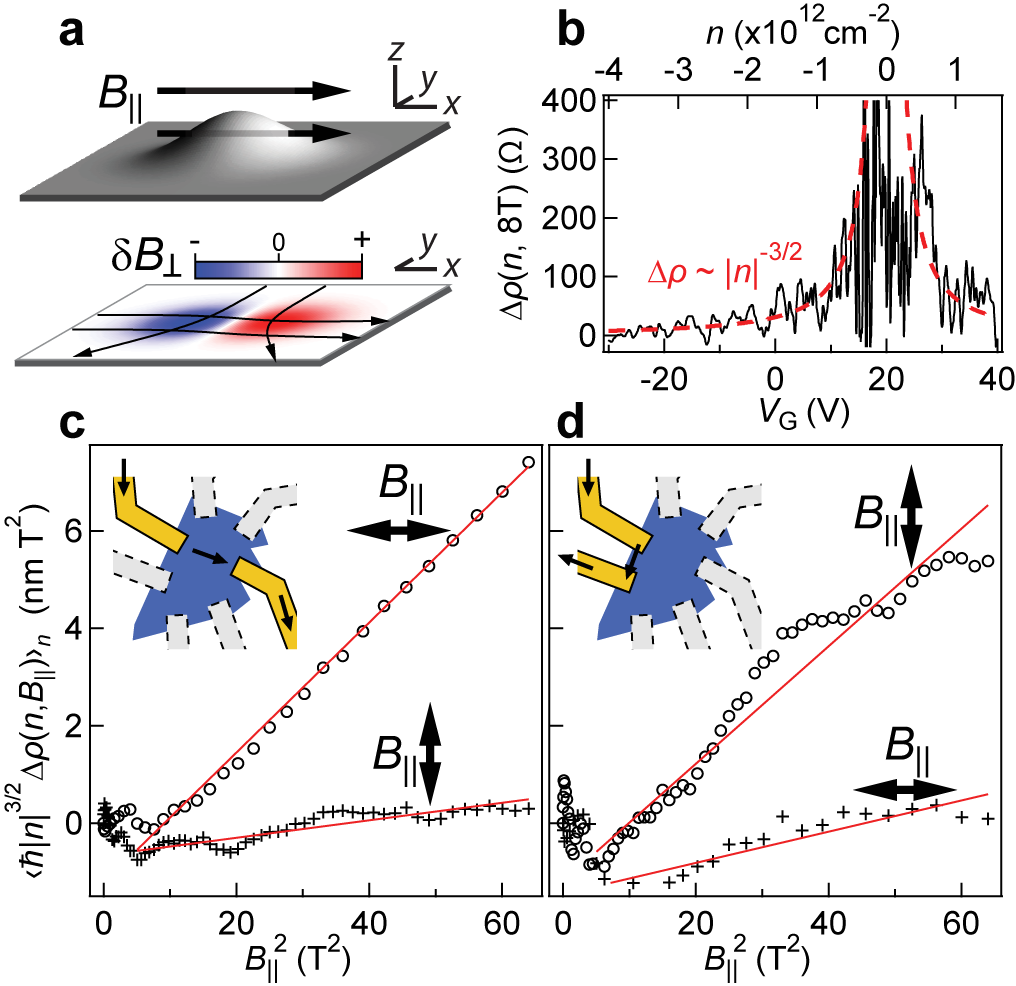}{245}{185}
\end{center}
\caption{
Anisotropic in-plane magnetoresistivity at $B_\perp > 50$~mT in flake A at 4K.
({\bf a}) Upper: a simulated symmetric bump in the graphene sheet, with uniform field $B_\parallel$ applied in the $\hat{x}$ direction.
Lower: The resulting surface-normal field, $\delta B_\perp$, is antisymmetric (positive on the right).
Simulated trajectories show how an electron's $\hat x$-velocity is randomized more quickly than its $\hat y$-velocity.
({\bf b}) Density-dependence of $\rho(8$~T$)-\rho(0$~T$)$ at $B_\perp=50$~mT, $\theta = 20^\circ$.
Predicted large-$n$ behaviour in \eqref{bpmr} for $\riph^2/\ripl=0.15$~nm shown as dashed curve.
({\bf c}) Averages $\langle\hbar |n|^{3/2} \Delta \rho(n,\vec{B}_\parallel)\rangle_n$ show the dependence on the magnitude and direction of $\vec{B}_\parallel$. The current path (depicted in inset) was measured at in-plane field orientations $\theta\approx 20^\circ$ ($\circ$) and $\theta\approx 70^\circ$ ($+$).
({\bf d}) Measurements on a different pair of electrodes ($\aspect \approx 1.6$) confirm that the anisotropy depends on the  anglular difference, $\theta$, between field and current.
} \label{fig3}
\end{figure}

The expected magnetoresistivity $\Delta\rho(B_\parallel) = \aspect[1/G(B_\parallel)-1/G(0)]$ for Gaussian ripples due to Lorentz forces from the RVP can be calculated by a Boltzmann approach, assuming $k_F\ripl \gg 1$ (high doping):
\begin{equation}
\Delta\rho(n, \theta, B_\parallel) = \frac{\sin^2 \theta + 3\cos^2 \theta}{4} \frac{1}{\hbar |n|^{3/2}} \frac{\riph^2}{\ripl} B_\parallel^2,
\label{bpmr}
\end{equation}
where $\theta$ is the angle of the current flow relative to $\vec{B}_\parallel$.
Equation \ref{bpmr} is derived in the Appendix, and is most conveniently measured as a change in resistivity (rather than conductance).
The threefold anisotropy predicted by Eq.~\eqref{bpmr} was observed in a GaAs 2DEG when in-plane field lines were rippled by nearby ferromagnets\cite{PhysRevB.70.193313}. In that case, as with rippled graphene, each individual magnetic ripple includes equal parts positive and negative magnetic flux, with a $\hat y$-oriented zero-field channel (for $\vec{B}_\parallel$ along $\hat x$; see Fig.~3a). In the strong-field limit, such channels might form magnetic waveguides\cite{PhysRevB.77.081403,PhysRevB.77.081404}, but we observed a weaker form: the $\hat x$-component of the velocity of a diffusive ensemble is randomized faster than the $\hat y$-component because $\hat y$-moving particles are deflected more (Fig.~3a).\cite{PhysRevB.70.193313}

The density dependence, $\Delta\rho(n) \propto |n|^{-3/2}$, predicted from Eq.~\ref{bpmr} can be seen in the experimental data (Fig.~3b),
on top of phase-coherent conductance (resistance) fluctuations due to the in-plane field\cite{spinucf}.
The resistivity saturated for $|n|\lesssim 10^{12}$~cm$^{-2}$, perhaps due to the breakdown of classical scattering when $\fermik \ripl \lesssim 1$. %$\Delta \rho_{\rm sat}=\frac{1}{2}(\pi B_\parallel \riph \ripl)^2/\hbar$. %%% Exact Gaussian result.
The fluctuations in Fig.~3b were minimized by measuring at 4K, and averaging the quantity $\hbar |n|^{3/2}\Delta\rho(n,\vec{B}_\parallel)$ over $n = (-3.5\ldots{-1})\times10^{12}$~cm$^{-2}$, allowing fits to Eq.~\ref{bpmr} (Fig.~3c,d).
It was confirmed that the magnitude of the effect did not change from 4K down to 40mK, though the amplitude of the fluctuations increased at low temperature as expected.

Flake A was measured with two current paths along $\theta \approx 20^\circ$ and $70^\circ$. The device was then re-cooled in a $90^\circ$-rotated orientation, to change $\theta \rightarrow \theta + 90^\circ$.
Fits of magnetoresistance curves to Eq.~\ref{bpmr} gave a range $\riph^2/\ripl \sim 0.05$--$0.2$~nm for flake A (Fig.~3cd).  The measured anisotropy $\Delta\rho(70^\circ)/\Delta\rho(20^\circ)$ was approximately $0.13\pm0.01$ for one current path (Fig.~3c) and $0.26\pm0.03$ for the other (Fig.~3d), whereas Eq.~\eqref{bpmr} predicts $0.44$.
In the single measurement of flake B, $\riph^2/\ripl \approx 0.02$--0.04~nm. 

Using the value $\riph^2\ripl \approx 1.7$~nm$^3$ from the analysis in Fig.~2 and the range of values of $\riph^2/\ripl$ reported above, $\riph = 0.6 \pm 0.1$~nm and $\ripl = 4 \pm 2$~nm are extracted for Gaussian-correlated rippling of Flake A (the spread in $\riph^2/\ripl$ is incorporated into uncertainties for $\riph$ and $\ripl$).
The values for $\riph$ and $\ripl$ from this work can be compared to values obtained from scanning probe measurements on these flakes, and to values reported in the literature for graphene flakes on SiO$_2$.
After the measurements described above, room temperature atomic force microscope (AFM) measurements were performed on flake A using an Asylum MFP3D-SA, after annealing the flake at 400$^\circ$C in a low pressure N$_2$/H$_2$ gas mixture to remove resist residues\cite{nanolettripples}. These measurements gave $\riph = 0.13\pm0.02$~nm and $\ripl= 10\pm5$~nm; limitations of vibration and drift prevented more accurate measurements.  AFM measurements on similar flakes in Ref.~\onlinecite{nanolettripples}  gave $\riph = 0.19$~nm and $\ripl=32$~nm, whereas scanning tunnelling microscope (STM) measurements in Ref.~\onlinecite{geringerripples} gave $\riph = 0.35$~nm, $\ripl\approx5$~nm\cite{geringerthanks}.
A consistent discrepancy is noted between scanning probe measurements and topographic parameters extracted from transport: the observed magneto-resistance in flakes A and B is  nearly a factor of 100 larger than would be expected from our own AFM measurements on flake A. Although the right-hand side of Eq.~\eqref{bpmr} is enhanced by a factor between one and two for realistic non-Gaussian correlations in ripples (see Appendix), this is far too small to explain the measured magneto-resistance.

The in-plane field couples to spins as well as to the motion of charges, leading to the possibility of spin-related effects on transport.\cite{spinucf} Three different types of spin-related effects are considered.  
First, Zeeman splitting of the Fermi level implies altered populations of spin-up and spin-down electrons\cite{spinucf} that screen charged impurities less efficiently, leading to a magnetoresistance\cite{hwang-bpmr} $\rho(B_\parallel)/\rho(0) \approx (2\times10^5$~cm$^{-2}\,$T$^{-2}) B_\parallel^2 /|n|$ for densities larger than the impurity broadening.  Quantitatively, however, this effect would be too small to be observed in our measurements.  Second, spin-flip scattering off magnetic impurities can lead to decoherence, adding to $\tau_\phi^{-1}$. This effect would be field dependent, as $B \gtrsim B_i$ will freeze impurities into their ground state\cite{bigbook}, disabling spin-flip dephasing at $B_i = k_{\rm B}T/g^* \mu_{\rm B} \sim 100$~mT. Such an effect would show up as a peak in $\tau_\phi^{-1}$ at zero field in data such as Fig.~2c. Based on the apparent absence of this peak, impurity-induced spin flips must be uncommon for electrons traversing our devices. Finally, magnetic impurities may also generate an RVP by their localized magnetic fields, but the strength of this RVP (and hence its contribution to $\tau_\phi$) would not change when the impurities align to $B_\parallel$.

Finally, we turn to an important analogy that can be drawn between the effects of an in-plane field and those of strain due to ripples. Since both the in-plane field and ripple strain generate random vector potentials that are directly correlated with ripple topography\cite{mathurbaranger,geimwl}, similar effects can be expected. In particular, strain is commonly believed to suppress weak anti-localization\cite{geimwl,tikhonenko,mccannwl}, but there is widespread disagreement about how to estimate the magnitude of the effect.  We argue that the suppression occurs through a short-range dephasing process much like that in Eq.~\ref{mbphi}, and that strain-induced dephasing can fully explain the observed suppression of anti-localization.

Strain RVP affects each valley oppositely with a magnitude that depends on $\ripl$ and $\riph$,\cite{geimwl} corresponding to a fictitious valley-dependent in-plane field of size $\hbar\riph/(e a_0 \ripl^2)$, where the lattice constant is $a_0 = 0.14$~nm.
The resulting valley-dependent dephasing affects $\tau_*^{-1}$,\cite{mccannwl} and by analogy with Eq.~2 we expect $\tau_*^{-1} \approx \speed \riph^4/(a_0^2 \ripl^3)$. Using this expression, the ripple dimensions extracted from our in-plane field measurements or from  STM measurements\cite{geringerripples} may fully explain the large $\tau_*^{-1} \sim 1$--10~ps$^{-1}$ observed in most graphene WL magnetoresistance experiments (see Table~\ref{fittable} and Refs.~\onlinecite{tikhonenko,2009NJPh...11i5021E,2010JPCM...22t5301C}).

This result contrasts with previous estimates of the intra-valley effect of ripple strain, which have assumed that the strain-induced effective magnetic field is truly random, with no requirement for flux compensation over multiple correlation lengths.\cite{geimwl,tikhonenko,PhysRevLett.98.136801} That assumption would imply long range correlations in the RVP.  The dephasing measurements in the first half of this paper show that ripples and resulting RVP correlations are instead short-range, as expected for adhesion to a polished wafer, and the analysis for $\tau_\phi$ in Eq.~2 should apply also to $\tau_*$.

In conclusion, transport measurements of graphene flakes in an in-plane magnetic field showed effects due to the magnetic flux threaded through the ripples. The use of an auxiliary out-of-plane field allowed two distinct effects to be separated: weak localization suppression (by dephasing) and overall anisotropic magnetoresistance (by Lorentz-force scattering). Besides allowing a determination of the ripples' typical height and length scale, these measurements provide insight as to how other short-range random vector potentials (such as that due to ripple strain) might affect transport in graphene.

%nanolettripples: D=0.190(+-5?)nm,   L=32(+-1)nm
%crommie group on charge puddles:   ripples are D=0.15(+-??)nm,   L unstated but multiscale.
%			http://arxiv.org/ftp/arxiv/papers/0902/0902.4793.pdf
%columbia people saw D~0.5nm, L~10nm (this is vague.)
%			http://www.pnas.org/content/104/22/9209.full#F2
%geringerripples: D = 0.32 nm, L looks like 20 nm?
%tikhonenko: D = 0.3nm, L = 10nm?  (looks like 15 nm)

\section*{Experimental methods}
Silicon wafers with a $\sim$300~nm wet thermal oxide were thinned to $\sim$260~nm oxide thickness by a CF$_4$/O$_2$ plasma, then cleaned
using a standard SC-1/SC-2 process. Within an hour of cleaning, flakes of graphene were deposited using the mechanical exfoliation technique\cite{novoselov} then located in an optical microscope. The Cr/Au electrodes were deposited in an e-beam lithography process using PMMA resist.
Immediately before the cooldowns for electrical measurements, devices were baked on a 125$^\circ$C hot plate.
Graphene flakes were confirmed to be single layers by quantum Hall effect measurements\cite{spinucf}; the backgate capacitance $ne/(\gate-V_0) = 8.0\pm 0.1 \times 10^{10}$~cm$^{-2}$$e$/V was calculated from resistivity minima in this data. % In all calculations of $n$, a voltage offset $V_0$ was used to cancel the average doping from contamination;
In each cooldown, the gate offset $V_0$ was determined by requiring the conductance minimum to occur at $n=0$. For Figs.~1 and 2, $V_0 = 1$~V in flake A\cite{spinucf} and $V_0 = 11.5$~V in flake B. The quality of flake A had decayed before the 4~K measurements of Fig.~3, so that $V_0$ varied from  13 to 23~V depending on cooldown and the current path.

Two magnets provided fields up to 120~mT oriented out-of-plane, and 12~T oriented roughly in-plane. A $\sim$1$^{\circ}$ misalignment of the in-plane axis was corrected by biasing the out-of-plane magnet: the reported $B_\perp$ values have been manually offset at each $B_\parallel$ so that the WL dip always occurs at $B_\perp=0$.
Conductance was measured by lock-in techniques with a 10~nA current bias (Fig.~1b). In flake B, conductance was measured
in a four-probe geometry (Fig.~1b), and we further reduced the effects of conductance fluctuations by averaging over two opposing sets of voltage probes with the same aspect ratio (Fig.~1b). In flake A, two-terminal conductance was calculated after subtracting a $3.2$~k$\Omega$ contact resistance. Aspect ratios of flake A were computed by a 2D Laplace equation solver, and aspect ratios of flake B were estimated from the device geometry. In either case, aspect ratios were complicated by the invasive nature of the contacts\cite{huardcontacts}. The different aspect ratios for hole and electron doping (Table \ref{fittable}) were determined by requiring the conductivity to be $n$-symmetric: $(\aspecti G)(n)=(\aspecti G)(-n)$.

\appendix*

\section{$B_\parallel$ anisotropic magnetoresistance}

We model diffusive particle motion by a probability distribution $f(\vec k)$, with evolution $\frac{{\rm d}}{{\rm d}t}[f(\vec k)] = D[S_0 + S_\parallel, f](\vec k) $, for $ D[S, f](\vec k) = \int \frac{{\rm d}^2 k'}{(2\pi)^2} S(\vec k, \vec k') (f(\vec k') - f(\vec k))$. The zero-field scattering matrix is $S_0$, assumed to have the elastic and isotropic form $S_0(\vec k, \vec k') = s_0(k, q)\delta(k'-k)$, where $\vec q = \vec k' - \vec k$. Transport occurs by modes $f_x(\theta) = \frac{1}{\sqrt{\pi}} \cos(\theta)$ and $f_y(\theta) = \frac{1}{\sqrt{\pi}} \sin(\theta)$ which are eigenmodes of $D[S_0,f_i] = -\tau_{ii}^{-1}f_i$ for all $k$. The zero-field transport time is then $\tau_{xx}=\tau_{yy}=\meanfreetime$, with $\tau_{xy}=0$.

The $B_\parallel$-induced scattering matrix is calculated by Fermi golden rule as
$S_\parallel(\vec k, \vec k') = \frac{2\pi V}{\hbar}| \langle \vec k' |  U(\vec r) | \vec k \rangle |^2 \delta(E_{k'}-E_k)$.
The potential is $U(\vec r) = - e v \vec A(\vec r)\cdot \vec \sigma$ for Dirac fermions with Pauli matrices $\vec \sigma$ operating on chirality. Given $B_\parallel$ oriented along $\hat x$, the vector potential is\cite{mathurbaranger} $\vec A(\vec r) = - B_\parallel h(\vec r) \hat y$ for rippled graphene with out-of-plane displacements denoted by $h(\vec r)$.
We consider just single-valley chiral plane waves
$|\vec k \rangle = \frac{1}{\sqrt{2V}} e^{i\vec k\cdot \vec r} [ e^{-i \theta_k / 2} ,  e^{i \theta_k / 2} ]^T$
under the assumption that $U$ does not mix valleys, yielding
\begin{equation}
S_\parallel(\vec k, \vec k') = \frac{2\pi e^2 v B_\parallel^2}{\hbar^2} c(\vec q)  \sin^2 \! \frac{\theta_{k'}+\theta_k}{2}
\delta(k-k'),
\label{smatrix}
\end{equation}
for height correlator $c(\vec r) = \langle h(\vec r_0) h(\vec r_0 + \vec r) \rangle$.
This scattering depends on the sum of $\theta_{k'}$ and $\theta_k$, and therefore may anisotropically break the degeneracy of the transport eigenmodes. As a perturbation of the evolution, this modifies the scattering rates as
$\delta\tau_{ij}^{-1} =  \int_{-\pi}^{\pi} {\rm d}\theta_k \, f_i(\theta_k) D [S_\parallel , f_j ](\theta_k)$, to first order.

Here we examine only the case of isotropic ripples $c(\vec q) = c(q)$, for which $\rho_{xy}, \rho_{yx}$ remain zero. After some calculation, the anisotropy is found to be exactly threefold:
$\Delta\rho_{xx} = \frac{3}{2}\rho_\parallel$, and $\Delta\rho_{yy} = \frac{1}{2}\rho_\parallel$, where $\rho_\parallel(k)$
may be written as a real-space integral of the height correlator:
\begin{equation}
\rho_\parallel(k) = \frac{ \pi B_\parallel^2}{\hbar} \int_0^{\infty} {\rm d}r\, r W(kr) c(r) ,
\label{bpmrr}
\end{equation}
for $W(z) = \int_{0}^{2\pi} {\rm d}\phi \, J_0(2z\sin \frac{\phi}{2})\sin^2 \frac{\phi}{2} $.
We have confirmed the threefold anisotropy of relaxation by simulating a classical charge moving in the $x$-$y$ plane with out-of-plane magnetic field $ B = - B_\parallel \frac{{\rm d}h}{{\rm d}x} \hat z$, for random isotropic $h(r)$. Simulation and experiment in Ref.~\onlinecite{PhysRevB.70.193313} also show $\Delta\rho_{xx} = 3\Delta\rho_{yy}$.

$W(z)$ is approximately constant for $z\ll 1$, so $\rho_\parallel(k)$ plateaus at a resistivity $\sim(\riph \ripl B_\parallel)^2/\hbar$ at low densities where $k \ll 1/\ripl$.
At higher density ($k \gg 1/\ripl$), the oscillatory $W(z)$ may be integrated out via a Hankel transform. This yields the classical result $\rho_\parallel \propto k^{-3}$ of Eq.~\ref{bpmr} with an effective $[\riph^2/\ripl]_{\rm eff} = -\int_0^{\infty} {\rm d} r\, \frac{1}{r} \frac{{\rm d}}{{\rm d}r}c(r)/\sqrt{\pi}$, which is exactly $\riph^2/\ripl$ if the ripples are Gaussian.

For realistic non-Gaussian correlation functions, and for $k \approx 1/\ripl$, Eq.~\ref{bpmr} may be used for $\riph$ and $\ripl$ statistically defined as in Ref.~\onlinecite{nanolettripples}, though with some error. For more accuracy, Eq.~\ref{bpmrr} should be computed directly.
Using the correlation function of Ref.~\onlinecite{nanolettripples}, Eq.~\ref{bpmrr} gives a magnetoresistance approximately 80\% larger than that expected from Eq.\ref{bpmr}, due to fractal scaling of ripples which causes $[\riph^2/\ripl]_{\rm eff} > \riph^2/\ripl$.
For the $\ripl \approx 5$~nm correlation function\cite{geringerthanks} of Ref.~\onlinecite{geringerripples}, $k \approx 1/\ripl$ for the range of $n$ examined in Fig.~3, and Eq.~\ref{bpmrr} gives the magnetoresistance a $k$-dependence slightly weaker than the classical $k^{-3}$. Averaging over $n$ in the method of Fig.~3 yields a result in agreement with Eq.~\ref{bpmr}, though strong disagreement arises for $\ripl < 4$~nm as $\rho_\parallel$ saturates to its maximum.

\begin{acknowledgments}
We acknowledge V.~F'alko, M.~Fuhrer, D.~Goldhaber-Gordon, C.~Marcus, and especially C.~Lewenkopf for helpful discussions.
Graphite provided by S.~Fain and D.~Cobden.  MBL acknowledges a PGS-D from NSERC; work funded by CIFAR, CFI, and NSERC.
\end{acknowledgments}

%\bibliography{bib}

\begin{thebibliography}{26}
\expandafter\ifx\csname natexlab\endcsname\relax\def\natexlab#1{#1}\fi
\expandafter\ifx\csname bibnamefont\endcsname\relax
  \def\bibnamefont#1{#1}\fi
\expandafter\ifx\csname bibfnamefont\endcsname\relax
  \def\bibfnamefont#1{#1}\fi
\expandafter\ifx\csname citenamefont\endcsname\relax
  \def\citenamefont#1{#1}\fi
\expandafter\ifx\csname url\endcsname\relax
  \def\url#1{\texttt{#1}}\fi
\expandafter\ifx\csname urlprefix\endcsname\relax\def\urlprefix{URL }\fi
\providecommand{\bibinfo}[2]{#2}
\providecommand{\eprint}[2][]{\url{#2}}

\bibitem[{\citenamefont{{Zhang} et~al.}(2005)\citenamefont{{Zhang}, {Tan},
  {Stormer}, and {Kim}}}]{kim}
\bibinfo{author}{\bibfnamefont{Y.}~\bibnamefont{{Zhang}}},
  \bibinfo{author}{\bibfnamefont{Y.-W.} \bibnamefont{{Tan}}},
  \bibinfo{author}{\bibfnamefont{H.~L.} \bibnamefont{{Stormer}}},
  \bibnamefont{and} \bibinfo{author}{\bibfnamefont{P.}~\bibnamefont{{Kim}}},
  \bibinfo{journal}{\nat} \textbf{\bibinfo{volume}{438}}, \bibinfo{pages}{201}
  (\bibinfo{year}{2005}).

\bibitem[{\citenamefont{Novoselov et~al.}(2004)\citenamefont{Novoselov, Geim,
  Morozov, Jiang, Zhang, Dubonos, Grigorieva, and Firsov}}]{novoselov}
\bibinfo{author}{\bibfnamefont{K.~S.} \bibnamefont{Novoselov}},
  \bibinfo{author}{\bibfnamefont{A.~K.} \bibnamefont{Geim}},
  \bibinfo{author}{\bibfnamefont{S.~V.} \bibnamefont{Morozov}},
  \bibinfo{author}{\bibfnamefont{D.}~\bibnamefont{Jiang}},
  \bibinfo{author}{\bibfnamefont{Y.}~\bibnamefont{Zhang}},
  \bibinfo{author}{\bibfnamefont{S.~V.} \bibnamefont{Dubonos}},
  \bibinfo{author}{\bibfnamefont{I.~V.} \bibnamefont{Grigorieva}},
  \bibnamefont{and} \bibinfo{author}{\bibfnamefont{A.~A.}
  \bibnamefont{Firsov}}, \bibinfo{journal}{Science}
  \textbf{\bibinfo{volume}{306}}, \bibinfo{pages}{666} (\bibinfo{year}{2004}).

\bibitem[{\citenamefont{{Meyer} et~al.}(2007)\citenamefont{{Meyer}, {Geim},
  {Katsnelson}, {Novoselov}, {Booth}, and {Roth}}}]{geimripples}
\bibinfo{author}{\bibfnamefont{J.~C.} \bibnamefont{{Meyer}}},
  \bibinfo{author}{\bibfnamefont{A.~K.} \bibnamefont{{Geim}}},
  \bibinfo{author}{\bibfnamefont{M.~I.} \bibnamefont{{Katsnelson}}},
  \bibinfo{author}{\bibfnamefont{K.~S.} \bibnamefont{{Novoselov}}},
  \bibinfo{author}{\bibfnamefont{T.~J.} \bibnamefont{{Booth}}},
  \bibnamefont{and} \bibinfo{author}{\bibfnamefont{S.}~\bibnamefont{{Roth}}},
  \bibinfo{journal}{\nat} \textbf{\bibinfo{volume}{446}}, \bibinfo{pages}{60}
  (\bibinfo{year}{2007}).

\bibitem[{\citenamefont{Ishigami et~al.}(2007)\citenamefont{Ishigami, Chen,
  Cullen, Fuhrer, and Williams}}]{nanolettripples}
\bibinfo{author}{\bibfnamefont{M.}~\bibnamefont{Ishigami}},
  \bibinfo{author}{\bibfnamefont{J.~H.} \bibnamefont{Chen}},
  \bibinfo{author}{\bibfnamefont{W.~G.} \bibnamefont{Cullen}},
  \bibinfo{author}{\bibfnamefont{M.~S.} \bibnamefont{Fuhrer}},
  \bibnamefont{and} \bibinfo{author}{\bibfnamefont{E.~D.}
  \bibnamefont{Williams}}, \bibinfo{journal}{Nano Lett.}
  \textbf{\bibinfo{volume}{7}}, \bibinfo{pages}{1643} (\bibinfo{year}{2007}).

\bibitem[{\citenamefont{Geringer et~al.}(2009)\citenamefont{Geringer, Liebmann,
  Echtermeyer, Runte, Schmidt, R\"{u}ckamp, Lemme, and
  Morgenstern}}]{geringerripples}
\bibinfo{author}{\bibfnamefont{V.}~\bibnamefont{Geringer}},
  \bibinfo{author}{\bibfnamefont{M.}~\bibnamefont{Liebmann}},
  \bibinfo{author}{\bibfnamefont{T.}~\bibnamefont{Echtermeyer}},
  \bibinfo{author}{\bibfnamefont{S.}~\bibnamefont{Runte}},
  \bibinfo{author}{\bibfnamefont{M.}~\bibnamefont{Schmidt}},
  \bibinfo{author}{\bibfnamefont{R.}~\bibnamefont{R\"{u}ckamp}},
  \bibinfo{author}{\bibfnamefont{M.~C.} \bibnamefont{Lemme}}, \bibnamefont{and}
  \bibinfo{author}{\bibfnamefont{M.}~\bibnamefont{Morgenstern}},
  \bibinfo{journal}{Phys. Rev. Lett.} \textbf{\bibinfo{volume}{102}},
  \bibinfo{eid}{076102} (\bibinfo{year}{2009}).

\bibitem[{\citenamefont{Tikhonenko et~al.}(2008)\citenamefont{Tikhonenko,
  Horsell, Gorbachev, and Savchenko}}]{tikhonenko}
\bibinfo{author}{\bibfnamefont{F.~V.} \bibnamefont{Tikhonenko}},
  \bibinfo{author}{\bibfnamefont{D.~W.} \bibnamefont{Horsell}},
  \bibinfo{author}{\bibfnamefont{R.~V.} \bibnamefont{Gorbachev}},
  \bibnamefont{and} \bibinfo{author}{\bibfnamefont{A.~K.}
  \bibnamefont{Savchenko}}, \bibinfo{journal}{Phys. Rev. Lett.}
  \textbf{\bibinfo{volume}{100}}, \bibinfo{eid}{056802} (\bibinfo{year}{2008}).

\bibitem[{\citenamefont{Morozov et~al.}(2006)\citenamefont{Morozov, Novoselov,
  Katsnelson, Schedin, Ponomarenko, Jiang, and Geim}}]{geimwl}
\bibinfo{author}{\bibfnamefont{S.~V.} \bibnamefont{Morozov}},
  \bibinfo{author}{\bibfnamefont{K.~S.} \bibnamefont{Novoselov}},
  \bibinfo{author}{\bibfnamefont{M.~I.} \bibnamefont{Katsnelson}},
  \bibinfo{author}{\bibfnamefont{F.}~\bibnamefont{Schedin}},
  \bibinfo{author}{\bibfnamefont{L.~A.} \bibnamefont{Ponomarenko}},
  \bibinfo{author}{\bibfnamefont{D.}~\bibnamefont{Jiang}}, \bibnamefont{and}
  \bibinfo{author}{\bibfnamefont{A.~K.} \bibnamefont{Geim}},
  \bibinfo{journal}{Phys. Rev. Lett.} \textbf{\bibinfo{volume}{97}},
  \bibinfo{eid}{016801} (\bibinfo{year}{2006}).

\bibitem[{\citenamefont{{Katsnelson} and {Geim}}(2008)}]{2008RSPTA.366..195K}
\bibinfo{author}{\bibfnamefont{M.~I.} \bibnamefont{{Katsnelson}}}
  \bibnamefont{and} \bibinfo{author}{\bibfnamefont{A.~K.}
  \bibnamefont{{Geim}}}, \bibinfo{journal}{Phil. Trans. R. Soc. A}
  \textbf{\bibinfo{volume}{366}}, \bibinfo{pages}{195} (\bibinfo{year}{2008}),
  \eprint{0706.2490}.

\bibitem[{\citenamefont{Mensz et~al.}(1987)\citenamefont{Mensz, Wheeler, Foxon,
  and Harris}}]{menszAlGaAs}
\bibinfo{author}{\bibfnamefont{P.~M.} \bibnamefont{Mensz}},
  \bibinfo{author}{\bibfnamefont{R.~G.} \bibnamefont{Wheeler}},
  \bibinfo{author}{\bibfnamefont{C.~T.} \bibnamefont{Foxon}}, \bibnamefont{and}
  \bibinfo{author}{\bibfnamefont{J.~J.} \bibnamefont{Harris}},
  \bibinfo{journal}{Appl. Phys. Lett.} \textbf{\bibinfo{volume}{50}},
  \bibinfo{pages}{603} (\bibinfo{year}{1987}).

\bibitem[{\citenamefont{{Sullivan} et~al.}(2004)\citenamefont{{Sullivan},
  {Kane}, and {Thompson}}}]{dephasSiP}
\bibinfo{author}{\bibfnamefont{D.~F.} \bibnamefont{{Sullivan}}},
  \bibinfo{author}{\bibfnamefont{B.~E.} \bibnamefont{{Kane}}},
  \bibnamefont{and} \bibinfo{author}{\bibfnamefont{P.~E.}
  \bibnamefont{{Thompson}}}, \bibinfo{journal}{Appl. Phys. Lett.}
  \textbf{\bibinfo{volume}{85}}, \bibinfo{pages}{6362} (\bibinfo{year}{2004}).

\bibitem[{\citenamefont{Minkov et~al.}(2004)\citenamefont{Minkov, Rut,
  Germanenko, Sherstobitov, Zvonkov, Shashkin, Khrykin, and
  Filatov}}]{dephasInGaAs}
\bibinfo{author}{\bibfnamefont{G.~M.} \bibnamefont{Minkov}},
  \bibinfo{author}{\bibfnamefont{O.~E.} \bibnamefont{Rut}},
  \bibinfo{author}{\bibfnamefont{A.~V.} \bibnamefont{Germanenko}},
  \bibinfo{author}{\bibfnamefont{A.~A.} \bibnamefont{Sherstobitov}},
  \bibinfo{author}{\bibfnamefont{B.~N.} \bibnamefont{Zvonkov}},
  \bibinfo{author}{\bibfnamefont{V.~I.} \bibnamefont{Shashkin}},
  \bibinfo{author}{\bibfnamefont{O.~I.} \bibnamefont{Khrykin}},
  \bibnamefont{and} \bibinfo{author}{\bibfnamefont{D.~O.}
  \bibnamefont{Filatov}}, \bibinfo{journal}{Phys. Rev. B}
  \textbf{\bibinfo{volume}{70}}, \bibinfo{pages}{035304}
  (\bibinfo{year}{2004}).

\bibitem[{\citenamefont{Mathur and Baranger}(2001)}]{mathurbaranger}
\bibinfo{author}{\bibfnamefont{H.}~\bibnamefont{Mathur}} \bibnamefont{and}
  \bibinfo{author}{\bibfnamefont{H.~U.} \bibnamefont{Baranger}},
  \bibinfo{journal}{Phys. Rev. B} \textbf{\bibinfo{volume}{64}},
  \bibinfo{pages}{235325} (\bibinfo{year}{2001}).

\bibitem[{\citenamefont{Geim et~al.}(1994)\citenamefont{Geim, Bending,
  Grigorieva, and Blamire}}]{geimrmf}
\bibinfo{author}{\bibfnamefont{A.~K.} \bibnamefont{Geim}},
  \bibinfo{author}{\bibfnamefont{S.~J.} \bibnamefont{Bending}},
  \bibinfo{author}{\bibfnamefont{I.~V.} \bibnamefont{Grigorieva}},
  \bibnamefont{and} \bibinfo{author}{\bibfnamefont{M.~G.}
  \bibnamefont{Blamire}}, \bibinfo{journal}{Phys. Rev. B}
  \textbf{\bibinfo{volume}{49}}, \bibinfo{pages}{5749} (\bibinfo{year}{1994}).

\bibitem[{\citenamefont{Rushforth et~al.}(2004)\citenamefont{Rushforth,
  Gallagher, Main, Neumann, Henini, Marrows, and Hickey}}]{PhysRevB.70.193313}
\bibinfo{author}{\bibfnamefont{A.~W.} \bibnamefont{Rushforth}},
  \bibinfo{author}{\bibfnamefont{B.~L.} \bibnamefont{Gallagher}},
  \bibinfo{author}{\bibfnamefont{P.~C.} \bibnamefont{Main}},
  \bibinfo{author}{\bibfnamefont{A.~C.} \bibnamefont{Neumann}},
  \bibinfo{author}{\bibfnamefont{M.}~\bibnamefont{Henini}},
  \bibinfo{author}{\bibfnamefont{C.~H.} \bibnamefont{Marrows}},
  \bibnamefont{and} \bibinfo{author}{\bibfnamefont{B.~J.}
  \bibnamefont{Hickey}}, \bibinfo{journal}{Phys. Rev. B}
  \textbf{\bibinfo{volume}{70}}, \bibinfo{pages}{193313}
  (\bibinfo{year}{2004}).

\bibitem[{\citenamefont{{Lundeberg} and {Folk}}(2009)}]{spinucf}
\bibinfo{author}{\bibfnamefont{M.~B.} \bibnamefont{{Lundeberg}}}
  \bibnamefont{and} \bibinfo{author}{\bibfnamefont{J.~A.}
  \bibnamefont{{Folk}}}, \bibinfo{journal}{Nature Phys.}
  \textbf{\bibinfo{volume}{5}}, \bibinfo{pages}{894} (\bibinfo{year}{2009}).

\bibitem[{\citenamefont{McCann et~al.}(2006)\citenamefont{McCann, Kechedzhi,
  Fal'ko, Suzuura, Ando, and Altshuler}}]{mccannwl}
\bibinfo{author}{\bibfnamefont{E.}~\bibnamefont{McCann}},
  \bibinfo{author}{\bibfnamefont{K.}~\bibnamefont{Kechedzhi}},
  \bibinfo{author}{\bibfnamefont{V.~I.} \bibnamefont{Fal'ko}},
  \bibinfo{author}{\bibfnamefont{H.}~\bibnamefont{Suzuura}},
  \bibinfo{author}{\bibfnamefont{T.}~\bibnamefont{Ando}}, \bibnamefont{and}
  \bibinfo{author}{\bibfnamefont{B.~L.} \bibnamefont{Altshuler}},
  \bibinfo{journal}{Phys. Rev. Lett.} \textbf{\bibinfo{volume}{97}},
  \bibinfo{eid}{146805} (\bibinfo{year}{2006}).

\bibitem[{\citenamefont{Wada et~al.}(2010)\citenamefont{Wada, Okuda, and
  Wakabayashi}}]{Wada20101138}
\bibinfo{author}{\bibfnamefont{S.}~\bibnamefont{Wada}},
  \bibinfo{author}{\bibfnamefont{N.}~\bibnamefont{Okuda}}, \bibnamefont{and}
  \bibinfo{author}{\bibfnamefont{J.}~\bibnamefont{Wakabayashi}},
  \bibinfo{journal}{Physica E: Low-dimensional Systems and Nanostructures}
  \textbf{\bibinfo{volume}{42}}, \bibinfo{pages}{1138 } (\bibinfo{year}{2010}),
  ISSN \bibinfo{issn}{1386-9477}, \bibinfo{note}{18th International Conference
  on Electron Properties of Two-Dimensional Systems}.

\bibitem[{\citenamefont{Oroszl\'any et~al.}(2008)\citenamefont{Oroszl\'any,
  Rakyta, Korm\'anyos, Lambert, and Cserti}}]{PhysRevB.77.081403}
\bibinfo{author}{\bibfnamefont{L.}~\bibnamefont{Oroszl\'any}},
  \bibinfo{author}{\bibfnamefont{P.}~\bibnamefont{Rakyta}},
  \bibinfo{author}{\bibfnamefont{A.}~\bibnamefont{Korm\'anyos}},
  \bibinfo{author}{\bibfnamefont{C.~J.} \bibnamefont{Lambert}},
  \bibnamefont{and} \bibinfo{author}{\bibfnamefont{J.}~\bibnamefont{Cserti}},
  \bibinfo{journal}{Phys. Rev. B} \textbf{\bibinfo{volume}{77}},
  \bibinfo{pages}{081403} (\bibinfo{year}{2008}).

\bibitem[{\citenamefont{Ghosh et~al.}(2008)\citenamefont{Ghosh, De~Martino,
  H\"ausler, Dell'Anna, and Egger}}]{PhysRevB.77.081404}
\bibinfo{author}{\bibfnamefont{T.~K.} \bibnamefont{Ghosh}},
  \bibinfo{author}{\bibfnamefont{A.}~\bibnamefont{De~Martino}},
  \bibinfo{author}{\bibfnamefont{W.}~\bibnamefont{H\"ausler}},
  \bibinfo{author}{\bibfnamefont{L.}~\bibnamefont{Dell'Anna}},
  \bibnamefont{and} \bibinfo{author}{\bibfnamefont{R.}~\bibnamefont{Egger}},
  \bibinfo{journal}{Phys. Rev. B} \textbf{\bibinfo{volume}{77}},
  \bibinfo{pages}{081404} (\bibinfo{year}{2008}).

\bibitem[{ger()}]{geringerthanks}
\bibinfo{note}{We thank the authors of Ref.~5 for providing raw STM data for
  statistical analysis, used to extract Z and R by the criteria given in Ref.
  4}.

\bibitem[{\citenamefont{Hwang and Sarma}(2009)}]{hwang-bpmr}
\bibinfo{author}{\bibfnamefont{E.~H.} \bibnamefont{Hwang}} \bibnamefont{and}
  \bibinfo{author}{\bibfnamefont{S.~D.} \bibnamefont{Sarma}},
  \bibinfo{journal}{Phys. Rev. B} \textbf{\bibinfo{volume}{80}},
  \bibinfo{eid}{075417} (\bibinfo{year}{2009}).

\bibitem[{\citenamefont{Meyer et~al.}(2002)\citenamefont{Meyer, Fal'ko, and
  Altshuler}}]{bigbook}
\bibinfo{author}{\bibfnamefont{J.~S.} \bibnamefont{Meyer}},
  \bibinfo{author}{\bibfnamefont{V.~I.} \bibnamefont{Fal'ko}},
  \bibnamefont{and} \bibinfo{author}{\bibfnamefont{B.~L.}
  \bibnamefont{Altshuler}}, \emph{\bibinfo{title}{NATO Science Series II, Vol.
  72}} (\bibinfo{publisher}{Kluwer Academic, Dordrecht}, \bibinfo{year}{2002}).

\bibitem[{\citenamefont{{Eroms} and {Weiss}}(2009)}]{2009NJPh...11i5021E}
\bibinfo{author}{\bibfnamefont{J.}~\bibnamefont{{Eroms}}} \bibnamefont{and}
  \bibinfo{author}{\bibfnamefont{D.}~\bibnamefont{{Weiss}}},
  \bibinfo{journal}{New Journal of Physics} \textbf{\bibinfo{volume}{11}},
  \bibinfo{pages}{095021} (\bibinfo{year}{2009}), \eprint{0901.0840}.

\bibitem[{\citenamefont{{Chen} et~al.}(2010)\citenamefont{{Chen}, {Bae},
  {Chialvo}, {Dirks}, {Bezryadin}, and {Mason}}}]{2010JPCM...22t5301C}
\bibinfo{author}{\bibfnamefont{Y.}~\bibnamefont{{Chen}}},
  \bibinfo{author}{\bibfnamefont{M.}~\bibnamefont{{Bae}}},
  \bibinfo{author}{\bibfnamefont{C.}~\bibnamefont{{Chialvo}}},
  \bibinfo{author}{\bibfnamefont{T.}~\bibnamefont{{Dirks}}},
  \bibinfo{author}{\bibfnamefont{A.}~\bibnamefont{{Bezryadin}}},
  \bibnamefont{and} \bibinfo{author}{\bibfnamefont{N.}~\bibnamefont{{Mason}}},
  \bibinfo{journal}{Journal of Physics Condensed Matter}
  \textbf{\bibinfo{volume}{22}}, \bibinfo{pages}{205301}
  (\bibinfo{year}{2010}), \eprint{0910.3737}.

\bibitem[{\citenamefont{Wu et~al.}(2007)\citenamefont{Wu, Li, Song, Berger, and
  de~Heer}}]{PhysRevLett.98.136801}
\bibinfo{author}{\bibfnamefont{X.}~\bibnamefont{Wu}},
  \bibinfo{author}{\bibfnamefont{X.}~\bibnamefont{Li}},
  \bibinfo{author}{\bibfnamefont{Z.}~\bibnamefont{Song}},
  \bibinfo{author}{\bibfnamefont{C.}~\bibnamefont{Berger}}, \bibnamefont{and}
  \bibinfo{author}{\bibfnamefont{W.~A.} \bibnamefont{de~Heer}},
  \bibinfo{journal}{Phys. Rev. Lett.} \textbf{\bibinfo{volume}{98}},
  \bibinfo{pages}{136801} (\bibinfo{year}{2007}).

\bibitem[{\citenamefont{Huard et~al.}(2008)\citenamefont{Huard, Stander,
  Sulpizio, and Goldhaber-Gordon}}]{huardcontacts}
\bibinfo{author}{\bibfnamefont{B.}~\bibnamefont{Huard}},
  \bibinfo{author}{\bibfnamefont{N.}~\bibnamefont{Stander}},
  \bibinfo{author}{\bibfnamefont{J.~A.} \bibnamefont{Sulpizio}},
  \bibnamefont{and}
  \bibinfo{author}{\bibfnamefont{D.}~\bibnamefont{Goldhaber-Gordon}},
  \bibinfo{journal}{Phys. Rev. B} \textbf{\bibinfo{volume}{78}},
  \bibinfo{eid}{121402} (\bibinfo{year}{2008}).

\end{thebibliography}

\clearpage

\par

\end{document}